\documentclass[12pt]{article}
\usepackage[margin=1in]{geometry}
\usepackage{amsmath, amssymb, amsthm}
\usepackage{graphicx}
\usepackage{hyperref}
\usepackage{setspace}
\usepackage{titlesec}
\usepackage{bbm}
\usepackage{dsfont}
\usepackage{float}
\usepackage{color}

\titleformat{\section}{\large\bfseries}{\thesection.}{0.5em}{}
\titleformat{\subsection}{\normalsize\bfseries}{\thesubsection}{0.5em}{}

\title{\Large\bf Differential ML with a Difference}

\author{
\begin{tabular}{c@{\hspace{3cm}}c}
Paul Glasserman & Siddharth Hemant Karmarkar \\
Columbia University & Columbia University \\
\texttt{pg20@columbia.edu} & \texttt{shk2195@columbia.edu}
\end{tabular}
}

\date{November 2025; revised April 2026}

\begin{document}

\maketitle

\begin{abstract}
Differential ML (Huge and Savine 2020) is a technique for training neural networks to provide fast approximations to complex simulation-based models for derivatives pricing and risk management. It uses price sensitivities calculated through pathwise adjoint differentiation to reduce pricing and hedging errors. However, for options with discontinuous payoffs, such as digital or barrier options, the pathwise sensitivities are biased, and incorporating them into the loss function can magnify errors. We consider alternative methods for estimating sensitivities and find that they can substantially reduce test errors in prices and in their sensitivities. Using differential labels calculated through the likelihood ratio method expands the scope of  Differential ML to discontinuous payoffs. A hybrid method incorporates gamma estimates as well as delta estimates, providing further regularization.
\end{abstract}


\baselineskip18pt

\section{Introduction}

Machine learning methods can provide fast approximations to complex derivatives pricing and risk management models. They can be trained based on historical or simulated data. In either case, the standard approach trains a model using a set of spot prices for the underlying assets (and any other market inputs) and the discounted payouts resulting from these initial values under one or more derivatives contracts. These payouts depend on the historical or simulated evolution of the market. In the language of machine learning, the
total payouts are labels; the goal of the machine learning approach is to
develop a fast approximation to the mapping from market inputs to the corresponding
labels. For general background on machine learning in finance, see Dixon et al.~\cite{dixon}; for an application to approximating volatilities rather than prices, see Horvath et al.~\cite{horvath}; for a convergence rate analysis, see Detering et al.~\cite{detering}.

Huge and Savine \cite{hs} extended this general approach by expanding
the set of labels used for training to include payout \emph{sensitivities} as
well as the payouts themselves. Related ideas have been used in the physics literature (Raissi et al. \cite{pinn}), but the financial setting has distinctive features.
Targeting sensitivities in training an approximation addresses two points. First, it should improve the model training process by providing addititional labels to target.
Second, it should improve the sensitivities (Greeks) calculated from the
machine learning approximation, and these risk measures are of paramount
importance in the financial setting. As Huge and Savine \cite{hs} put it,
``the derivatives of a good approximation are not always a good approximation
of the derivatives.'' Their differential machine learning method (DML) seeks to
overcome this limitation.

For their differential labels (the payout sensitivities used in training), Huge and Savine \cite{hs} use \emph{pathwise} derivatives. Numerical results confirm that, when
applicable, targeting these differential labels reduces pricing errors and yields more accurate delta approximations than standard ML using only payout labels. However, the pathwise method yields biased estimates for derivative contracts with digital or barrier features. (See, e.g., the discussion in Section 7.2.2 of Glasserman \cite{pgbook}.) Targeting biased differential labels can severely  magnify errors and undermine the objective of DML.

To address this issue, we expand the set of differential labels for DML beyond pathwise sensitivities through the \emph{likelihood ratio method} (LRM) and a
hybrid method. LRM is applicable even with discontinuous payouts, such as those
resulting from digital or barrier features. Our hybrid method allows extending DML to include second-order (gamma) sensitivities. We validate these ideas through numerical examples. We also discuss limitations of the LRM method.

Huge and Savine \cite{hs} note in passing that the pathwise method may require ``appropriate smoothing.'' Smoothing introduces its own bias, and may be complicated to implement with path-dependent discontinuities. We will revisit smoothing in Section~\ref{s:smooth}. But the main objective of this article is to show that the scope of DML can be expanded through LRM-based differential labels, complementing the use of pathwise estimates, with or without smoothing.

Section~\ref{s:DML} provides background on standard and differential machine learning.
Section~\ref{s:disc} illustrates the problem of discontinuous payouts, explains the use of LRM, and tests its performance on examples. Section~\ref{s:gamma} extends  the approach to include gamma-regularization.
Our implementation is available at
https://github.com/diff-ml-with-a-difference/options-pricing.

\section{Differential Machine Learning}
\label{s:DML}

\subsection{Problem Setup and Standard Learning}

We generally follow the setting and notation of Huge and Savine \cite{hs}.
We consider a model specified through a Monte Carlo simulation algorithm, represented by a function $h$.
The algorithm takes as input $x$, a vector of spot prices (and possibly other market variables),
it applies stochastic inputs $\xi$, and returns the discounted payout
$h(x,\xi)$. This payout could apply to a portfolio of path-dependent options on multiple underlying assets, priced through a complex model with stochastic volatility and other features captured through the stochastic inputs $\xi$ and the function $h$.
The function $h$ embodies the simulation algorithm and is typically defined only implicitly through a combination of a large number of simpler operations.

The pricing function for this model is given by
\begin{equation}
f(x) = \mathbb{E}[h(x,\xi)],
\end{equation}
the expectation taken over the stochastic inputs $\xi$. In principle, the
pricing function can be approximated by averaging over repeated Monte Carlo samples, but this may be prohibitively time-consuming for real-time use.
In such settings, developing a faster approximation that can be trained off-line
becomes attractive.

The standard training data consists of a set of pairs $(x^{(i)},y^{(i)})$, $i=1,\dots,m$, where the $x^{(i)}$ are spot-price vectors, and the labels $y^{(i)}$
are the corresponding payouts $y^{(i)} = h(x^{(i)},\xi^{(i)})$. The labels
are generated by running the Monte Carlo simulation from $x^{(i)}$ with
stochastic inputs $\xi^{(i)}$, which can be done off-line.

A conventional neural network defines a family of functions $\hat{f}(x; w)$ 
with parameters $w$. These parameters are the weights used in the
neural network. They are chosen to approximate
the true pricing function $f$ by minimizing the empirical mean-squared error (MSE),
\begin{equation}
\mathcal{L}_{\text{val}}(w) = \frac{1}{m}\sum_{i=1}^m
\big(\hat{f}(x^{(i)}; w) - y^{(i)}\big)^2,
\label{MSE}
\end{equation}
possibly with a regularization penalty.
Once trained, the approximation $\hat{f}(x;w)$ can be used for fast
evaluation at new spot prices $x$. However, even if
the error in (\ref{MSE}) is small, there is no guarantee that 
sensitivities calculated from $\hat{f}$ will be close to the sensitivities
for the true pricing function $f$.

To help fix ideas, consider the simple example of Black-Scholes option pricing.
The terminal value of the underlying asset is given by
\begin{equation}
S_T = x e^{(r-\sigma^2) T + \sigma \sqrt{T}\xi},
\label{gbm}
\end{equation}
with spot price $x$, and $\xi$ having a standard normal distribution.
The discounted payoff of a standard call option is $h(x,\xi) = e^{-rT}(S_T-K)^+$.
The training data in this case would take the form $(x^{(i)},y^{(i)})$,
for some spot prices $x^{(i)}$ and corresponding payoffs $y^{(i)} = h(x^{(i)},\xi^{(i)})$. With enough training data and a sufficiently rich network,
the minimizing $\hat{f}$ in (\ref{MSE}) will approximate the Black-Scholes
formula, though the formula itself is never used in the training process.
In practice, such an approach is used with models that do not admit closed-form
prices.

In our implementation, we generate $k$ payoffs $y^{(i,1)},\dots,y^{(i,k)}$ from each
input $x^{(i)}$ and take the label $y^{(i)}$ to be the average
\begin{equation}
y^{(i)} = \frac{1}{k}\sum_{j=1}^k y^{(i,j)}.
\label{kavg}
\end{equation}
This reduces the variability in the labels, but it also reduces the number of samples $m$ we
can process in a fixed amount of computing time. 
Our numerical results use $k=10$ unless otherwise indicated.

\subsection{Training Sensitivities with Prices}

In DML, each sample $(x^{(i)}, y^{(i)})$ is augmented with a differential label
$\delta^{(i)} = \nabla_xy^{(i)}$.
The training objective becomes
\begin{equation}
\mathcal{L}_{\text{DML}}(w) =
\frac{1}{m}\sum_{i=1}^m
\big(\hat{f}(x^{(i)}; w) - y^{(i)}\big)^2
+ \lambda \cdot
\frac{1}{m}\sum_{i=1}^m
\big\| \nabla_x \hat{f}(x^{(i)}; w)
- \delta^{(i)} \big\|^2,
\label{eq:DMLloss}
\end{equation}
where $\lambda$ balances the value and gradient terms.  

The new terms in (\ref{eq:DMLloss}) require careful explanation. We will discuss
$\nabla_x \hat{f}(x^{(i)}; w)$ first, and then turn to $\delta^{(i)}$.

Related ideas have been developed in the physics literature under the name of ``physics-informed neural networks'' (Raissi et al. \cite{pinn}). In that setting, physical principles impose a condition on the target function $f$ of the form $D(f)=0$, for some differential operator $D$. In place of the second term in (\ref{eq:DMLloss}),
the loss function penalizes deviations $\|D(\hat{f}(x^{(i)};w))\|^2$.
Our focus is on the proper choice of differential labels $\delta^{(i)}$ in (\ref{eq:DMLloss}) whereas in the physics-informed setting the corresponding labels are simply zero.

\subsection{Differentiation through Backpropagation}

The gradient $\nabla_x \hat{f}(x; w)$ in (\ref{eq:DMLloss}) is the sensitivity of the approximating
function $\hat{f}$ with respect to the spot prices $x$. When $\hat{f}$
is defined through a feedforward neural network, this gradient
can be calculated very efficiently through backpropagation (see, e.g., Section 6.5 of Goodfellow et al. \cite{gbc}), which is a
special case of automatic adjoint differentiation (AAD). 

This step is illustrated nicely in Figure 1 of Huge and Savine \cite{hs}.
Their mathematical specification goes as follows.
Consider a feedforward network with $L$ layers defined by the equations
\begin{align}
z_0 &= x, \label{l1} \\
z_l &= g_{l-1}(z_{l-1}) W_l + b_l, \quad l = 1,\dots,L, \label{l2} \\
\hat{f}(x; w) &= z_L. \label{l3}
\end{align}
Here, (\ref{l1}) assigns the vector of inputs $x$ to the initial nodes; step (\ref{l2}) applies
the elementwise activiation functions $g_{l-1}$, weight matrices $W_l$, and biases $b_l$ to calculate layer $l$ from layer $l-1$; and (\ref{l3}) outputs that the final result.
  
Backpropagation differentiates $\hat{f}(x; w)$ with respect to the inputs in the reverse direction by setting
\begin{align}
\bar{z}_L &= 1, \label{aad1} \\
\bar{z}_{l-1} &= (\bar{z}_l W_l^\top) \circ g'_{l-1}(z_{l-1}), \quad l = L, \dots, 1, \label{aad2}\\
\nabla_x \hat{f}(x; w) &= \bar{z}_0, \label{aad3}
\end{align}
where $\circ$ denotes elementwise multiplication, and $\bar{z}_l$ is the
derivative of the output $\hat{f}(x;w)$ with respect to $z_l$.
This backward recursion mirrors the forward propagation (\ref{l1})--(\ref{l3}), producing sensitivities with a cost comparable to an additional forward pass, independent of input dimension.

By concatenating the forward and backward computations, Huge and Savine \cite{hs} obtain a \emph{twin network}: one half predicts the value $\hat{f}(x)$, the other its derivatives $\nabla_x \hat{f}(x)$.  
The two share the same parameters and structure, and the composite graph supports joint optimization using the loss in \eqref{eq:DMLloss}.  

\subsection{Differential Labels}

Backpropagation computes the gradients $\nabla_x \hat{f}(x; w)$ in (\ref{eq:DMLloss}). It remains to consider the differential labels $\delta^{(i)}$
in (\ref{eq:DMLloss}). 

Huge and Savine \cite{hs} use pathwise differentiation to calculate the
labels
\begin{equation}
\delta^{(i)} = \nabla_x h(x^{(i)},\xi^{(i)}),
\label{pwdelta}
\end{equation}
where, as before, $h$ is the function implicitly defined by the steps of a Monte Carlo algorithm. Equation (\ref{pwdelta}) is called a pathwise derivative because it
is the derivative with the path of stochastic inputs $\xi$ held fixed. 

In the Black-Scholes example (\ref{gbm}), we get
\begin{equation}
\delta = e^{-rT}\partial_x (S_T-K)^+ = \mathds{1}\{S_T>K\}\frac{e^{-rT}S_T}{x},
\label{pwcall}
\end{equation}
where we used $\partial_xS_T = S_T/x$ from (\ref{gbm}).
The payoff fails to be differentiable at $S_T=K$, but this outcome has probability zero, so the pathwise derivative is well-defined. It will often happen that the simulation
algorithm $h$ fails to be differentiable at certain points, but as long as these points have probability zero, the pathwise derivative is well-defined.

As in prior work, Huge and Savine \cite{hs} find that the calculation of pathwise sensitivities can be greatly accelerated through AAD. However, it is important to emphasize that there are two entirely separate applications of AAD in Huge and Savine \cite{hs}: one computes $\nabla_x \hat{f}(x;w)$ through backpropagation in the neural network; the other computes pathwise derivatives in generating training data as input to the neural network. One can adopt the first application without the second --- pathwise derivatives (whether or not calculated through AAD) are not the only source of differential labels. We will argue that for discontinuous payoffs we need different labels.

%
%
%
%
%
%

\section{Dealing with Discontinuities}
\label{s:disc}
To understand what we want from the differential lables $\delta^{(i)}$, it
is helpful to revisit the standard labels $y^{(i)}$ and the standard loss function (\ref{MSE}). By the definition of the pricing function $f$, the payout labels have the property $f(x^{(i)}) = \mathbb{E}[h(x^{(i)},\xi^{(i)})]=\mathbb{E}[y^{(i)}]$;
the discounted payouts are unbiased estimates of the price. This property
ensures that minimizing (\ref{MSE}) drives the approximation $\hat{f}$ close to the true pricing function $f$.

In (\ref{eq:DMLloss}), we similarly want to drive $\nabla_x\hat{f}$ close
to $\nabla_x f$. To this end, we want
\begin{equation}
\nabla_x f(x^{(i)}) = \mathbb{E}[\delta^{(i)}];
\label{unbiased}
\end{equation}
in other words, \emph{we want the differential label to be an unbiased estimate of the true sensitivity}. In the case of pathwise derivatives, (\ref{unbiased}) requires
\begin{equation}
\nabla_x \mathbb{E}[h(x,\xi)] = \mathbb{E}[\nabla_x h(x,\xi)].
\label{unbiased2}
\end{equation}
When this condition fails, targeting a biased differential label in (\ref{eq:DMLloss})
steers $\nabla_x \hat{f}$ away from the true value $\nabla_x f$, undermining the objective of DML.

Conditions for (\ref{unbiased2}) have been widely studied; see the discussion in Section 7.2.2 of Glasserman \cite{pgbook}. A sufficient condition is that $h$ be Lipschitz in $x$. A simple rule of thumb in practice is that (\ref{unbiased2}) typically holds when $h$ is continuous in $x$ and fails otherwise.

\subsection{Digital Options}

Digital options provide a stark example of this phenomenon.
In the Black-Scholes model (\ref{gbm}), consider the payoff function
\begin{equation}
h(x,\xi) = e^{-rT}\mathds{1}_{\{S_T > K\}}.
\end{equation}
The event $S_T=K$ has probability zero, so the pathwise derivative of $h$ is well-defined. However, the payoff is piecewise constant, so the pathwise derivative is identically equal to zero, resulting in  $\delta^{(i)}\equiv 0$,
even though the expected payoff
\begin{equation}
f(x) = e^{-rT}\Phi\left(\frac{\log (x/K) + rT}{\sigma\sqrt{T}}-\frac{\sigma\sqrt{T}}{2}\right)
\label{fdigital}
\end{equation}
is strictly increasing in $x$, where $\Phi$ is the standard normal cumulative 
distribution function.

We illustrate the consequences of these biased labels through a numerical example, taking $r=0$, $T=1/3$, and $\sigma=0.20$. 
We give equal weight to the price and delta components of the loss function (\ref{eq:DMLloss}).
All of our numerical results are based on the code provided by Huge and Savine \cite{hs}, with relatively minor changes.%
\footnote{As in Huge and Savine \cite{hs}, each network has four hidden layers of twenty units each and uses softplus activations. All models are optimized with Adam and a cosine-decay learning rate
schedule between $10^{-3}$ and $10^{-6}$, using batch sizes of 256–512 for stability.}

\begin{figure}[H]
    \centering
        \includegraphics[trim=0.9in 7.5in 0in 0.8in,clip,width=0.95\textwidth]{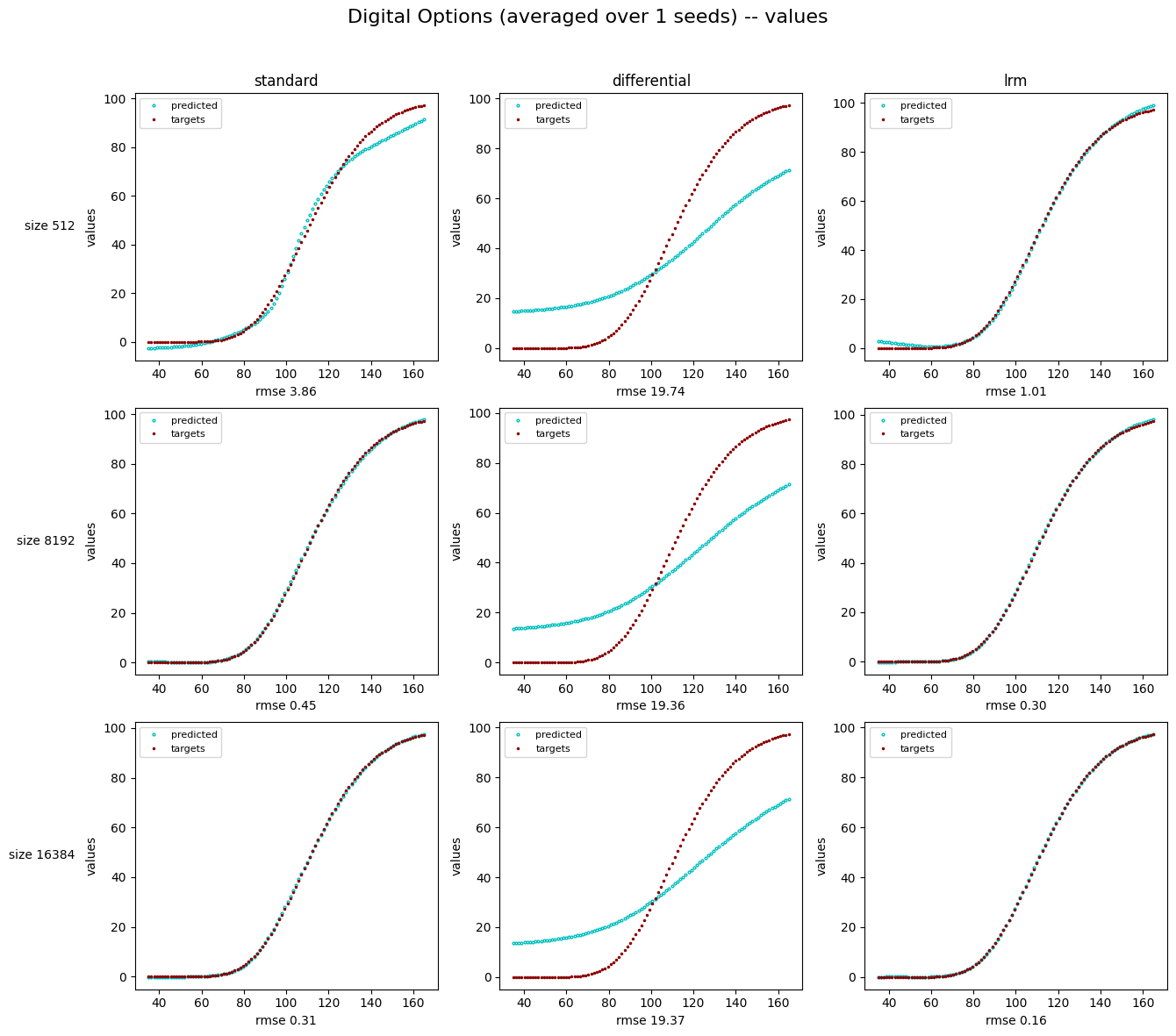}
    \caption{Digital options — predicted vs. target prices under standard, pathwise differential, and LRM differential training, using a sample size of 512 training labels.}
    \label{fig:digital_values}
\end{figure}

The left panel of Figure~\ref{fig:digital_values} shows results using standard ML based on the loss function (\ref{MSE}) without differential labels. We generate training data by generating $m=512$ values for the underlying asset and generating $k=10$ terminal values $S_T$ from each of these, as in (\ref{kavg}). After training, we test the approximation by evaluating it at levels of the underlying ranging from 40 to 160. We compare the approximation with the Black-Scholes digital prices and calculate the root mean squared error (RMSE) across these differences. 

In the left panel of the figure, the predicted values (from the neural network) are close to the correct values from (\ref{fdigital}). In the center panel, we repeat the same procedure but now using the DML objective (\ref{eq:DMLloss}) with the pathwise differential labels, which are identically zero for the digital payoff. The result is a significant failure. The predicted prices are far too high at low spot prices and far too low at high spot prices. 
The RMSE, which was 3.86 using standard ML, is now 19.71. As expected, targeting biased estimates of sensitivities steers the approximation far from the true target.
Because the pathwise derivativies are identically zero, by targeting these we produce
an approximation that is too flat.

To address this problem, we apply the \textit{likelihood ratio method} (LRM), which replaces pathwise derivatives with log-likelihood sensitivities.  
Whereas the pathwise method treats the input $x$ as an argument of the function $h$, LRM treats it as a parameter of the distribution of the underlying asset. 
(For general background on LRM, see Chapter VII of Asmussen and Glynn \cite{ag} or Section 7.3 of Glasserman \cite{pgbook}.)

Consider, for example, the Black-Scholes setting (\ref{gbm}) with spot price $S_0=x$. 
For any  discounted payoff $\pi(S_T)$, we have
$$
f(x) = \mathbb{E}[\pi(S_T)] = \int_0^{\infty} \pi(s)p(s;x)\, ds,
$$
where $p(s;x)$ is the lognormal density of $S_T$ when $S_0=x$,
\begin{equation}
p(s;x) = \frac{1}{x\sigma\sqrt{T}}\varphi\left(\frac{\log (s/x) - rT}{\sigma\sqrt{T}}+\frac{\sigma\sqrt{T}}{2}\right),
\label{lognor}
\end{equation}
with $\varphi$ the standard normal density.
LRM uses the identity
$$
f'(x) = \int_0^{\infty}[\pi(x)\nabla_x\log p(s;x)] p(x;s)\, ds
= \mathbb{E}[\pi(S_T)\nabla_x\log p(S_T;x)],
$$
to arrive at the unbiased estimator
\begin{equation}
\pi(S_T)\nabla_x\log p(S_T;x).
\label{pilrm}
\end{equation}
In the case of (\ref{lognor}), the \emph{score function} $\nabla_x\log p(S_T;x)$
evaluates to
\begin{equation}
\nabla_{x}\log p(S_T;x)=\frac{\xi}{x\sigma\sqrt{T}},
\label{score}
\end{equation}
with $\xi$ the same normal random variable used to generate the asset price $S_T$
in (\ref{gbm}).
This approach holds for any payoff $\pi(\cdot)$.
In the case of a digital option, the unbiased LRM delta estimator reads
\begin{equation}
\widehat{\Delta}_{\mathrm{LRM}}
=e^{-rT}\,\mathds{1}_{\{S_T>K\}}\,
\frac{\xi}{x\sigma\sqrt{T}}.
\label{eq:lrm_digital}
\end{equation}
We may therefore use as unbiased differential labels
\begin{equation}
\delta^{(i)} = e^{-rT}\,\mathds{1}_{\{S^{(i)}_T>K\}}\,
\frac{\xi^{(i)}}{x^{(i)}\sigma\sqrt{T}},
\label{lrmdiffs}
\end{equation}
in place of pathwise derivatives, which are biased for digital options.

It is worth emphasizing that the only change required in differential ML to implement this alternative is to replace the pathwise labels $\delta^{(i)}$ (which are identically zero for a digital option) with the LRM labels (\ref{lrmdiffs}) in (\ref{eq:DMLloss}). All other steps in the training and application process are unchanged. In particular, derivatives of the resulting approximation $\hat{f}(x;w)$ can be evaluated through backpropagation exactly as before.

In the right panel of Figure~\ref{fig:digital_values} we see that the LRM differential labels
yield a near-perfect approximation, in contrast to the pathwise labels in the middle panel.
The LRM approximation has an RMSE of 1.01, which is nearly a 20-fold reduction compared to pathwise DML and nearly a 4-fold reduction compared to standard ML. 
These correspond to MSE ratios of 400 and 16, respectively.
For a fixed method, MSE is generally inversely proportional to sample size, so the MSE ratios measure roughly how much more training data the other methods would need to match the accuracy of the LRM method.

Figure~\ref{fig:digital_values} compares prices $\hat{f}(x;w)$. As already noted, regardless of the method used to train the parameters $w$, the approximation can be differentiated using  (\ref{aad1})--(\ref{aad3}).  We therefore evaluate the error in $\partial_x\hat{f}(x;w)$
as an approximation to $\partial_x f(x)$. We call this the Delta RMSE to contrast it with the Price RMSE. The Delta RMSE using LRM is 34 times smaller than that for pathwise DML.

\subsection{Variance of LRM and Computational Considerations}
\label{s:variance}

It has been widely observed that, when applicable, the pathwise method generally produces
lower-variance estimates than LRM. In the absence of discontinuities, this favors the pathwise method.
There are two problem settings in which LRM estimates
often have high variance: very short maturities and a large number of time steps.
The potential problem at short maturities can be seen in (\ref{score}),
where the variance of the score function is $O(1/T)$.
We will examine the impact of increasing the number of time steps in Section~\ref{s:barrier}
and explain there why it does not lead to large variance in our setting.

The variance contributed by the score can be mitigated by using the score itself as a
control variate. By construction, the score always has mean zero and is thus
readily available as a control. It can be particularly effective when the score is the
primary source of variance in an LRM estimator.

Variance can be reduced with any derivative labels by
increasing the number of subpaths $k$ in (\ref{kavg}). This replaces a single
derivative label with the average over multiple draws.
To hold fixed the total computing
time, this means reducing the number of price scenarios $x^{(i)}$ used in training. 
Optimizing the relative values of the number of inputs $m$ and subpaths $k$ requires
some experimentation (using either pathwise or LRM labels). 
Recall that we think of the model training as taking place offline to allow fast
execution for online pricing and hedging, making such an approach feasible.

Indeed, it should be emphasized that any difficulties resulting from
large variance can be offset by generating more training data. In contrast, the bias
from applying the pathwise method to a discontinuous payoff will not be reduced by
increasing the sample size.

A comparison of estimator variances should be paired with a comparison of computing times. Across all the examples we have tested, the computing times using pathwise and LRM estimators are nearly identical. Moreover, the time to generate labels using either method is small compared with the time required to train the neural network.

Larger variance in the derivative labels will tend to favor a smaller $\lambda$
in the loss function (\ref{eq:DMLloss}). In all the experiments in this article, we
fix $\lambda =1$, giving equal weight to the payoff and sensitivity labels.
This is the default in the implementation of Huge and Savine \cite{hs} for
options on a single underlying.
In all cases, we have also explored systematically varying $\lambda$ from 0.1 to 10 to
see how the RMSE varies; this comparison is illustrated for the digital call example
in Table~\ref{t:lambda}. 
Some examples benefit from smaller $\lambda$ and some from
larger $\lambda$, but simply fixing $\lambda=1$ is a robust choice. One could alternatively
train models with small sample sizes at a few different values of $\lambda$, pick the
value with the smallest RMSE, and use that value on the full training data. This applies using either pathwise or LRM labels.

\begin{table}[H]
\begin{center}
\begin{tabular}{c|ccccccc}
$m/\lambda$	&	0.1	&	0.25	&	0.5	&	1	&	2	&	5	&	10 \\ \hline
1024	&	0.012	&	0.010	&	0.009	&	0.010	&	0.012	&	0.015	&	0.018 \\
8192	&	0.005	&	0.004	&	0.004	&	0.004	&	0.005	&	0.006	&	0.007 \\ \hline
\end{tabular}
\end{center}
\caption{Price RMSE for digital calls using LRM at different sample size $m$
and loss parameters $\lambda$.}
\label{t:lambda}
\end{table}

\subsection{Basket Digitals}

We turn next to a high-dimensional example. Following Huge and Savine \cite{hs}, we consider a basket option in the Bachelier model. This setting provides a convenient test case because it allows us to compare neural network approximations against closed-form prices regardless of dimension (since the Gaussian property is preserved by linear combinations). Whereas Huge and Savine \cite{hs} considered a standard basket call option, we consider a digital payouff.

For a basket with $d$ assets and weights $w_i$, the payoff is
\begin{equation}
\pi(S_{T,1},\ldots,S_{T,d}) = 
e^{-rT}\mathds{1}_{\left\{\sum_{i=1}^{d} w_i S_{T,i} > K\right\}}.
\end{equation}
The underlying assets are uncorrelated and evolve according to
\begin{equation}
S_{T,i} = x_i +  \sigma \sqrt{T}\,\xi_i,\; i=1,\dots,d, 
\qquad \xi \sim \mathcal{N}(0, I).
\end{equation}
The discontinuity now occurs on a $(d-1)$-dimensional hyperplane separating in- and out-of-the-money regions. This hyperplane has probability zero under the Gaussian distribution, so pathwise derivatives $\partial \pi/\partial S_{T,i}$ are well-defined. However,  they are identically zero.

The likelihood-ratio gradient estimator remains well-defined in this setting.  
Differentiating the log-density with respect to the initial spot $x_i$ gives the per-asset score
\begin{equation}
\nabla_{x_i} \log p(S_T; x) = 
\frac{\xi_i}{\sigma_i \sqrt{T}},
\end{equation}
leading to the componentwise LRM delta estimator
\begin{equation}
\widehat{\Delta}_{\text{LRM}}^{(i)} 
= e^{-rT}\, \mathds{1}_{\{\sum_j w_j S_{T,j} > K\}}\,
\frac{\xi_i}{\sigma_i \sqrt{T}}.
\label{eq:lrm_basket}
\end{equation}

Following Huge and Savine \cite{hs}, we estimate the average delta across assets and thus use the average of (\ref{eq:lrm_basket}) over $i=1,\dots,d$ as the differential label. With an equally weighted 20-asset basket, LRM achieves a 7-fold reduction in Price RMSE and 6-fold reduction in Delta RMSE compared to pathwise DML.

 \subsection{Smoothing}
 \label{s:smooth}
 
 Huge and Savine \cite{hs} note in passing that application of pathwise differentiation may require ``appropriate smoothing.'' Smoothing can be helpful in reducing bias, but it does not eliminate bias. Moreoever, the proper degree of smoothing in specific cases is far from obvious, and a suitable approach to smoothing may be unclear in multidimensional problems.
 
Consider again the simple setting of a digital payoff. The step function at the strike $K$ can be replaced with a ramp over the interval from $K-\epsilon/2$ to $K+\epsilon/2$. The pathwise method can be applied to this piecewise linear, continuous payoff.
 But how should $\epsilon$ be chosen? Taking $\epsilon$ too large increases bias; taking it too small increases variance.
 
\begin{figure}[H]
    \centering
        \includegraphics[trim=0.0in 0.0in 8in 0.52in,clip,width=0.55\textwidth]{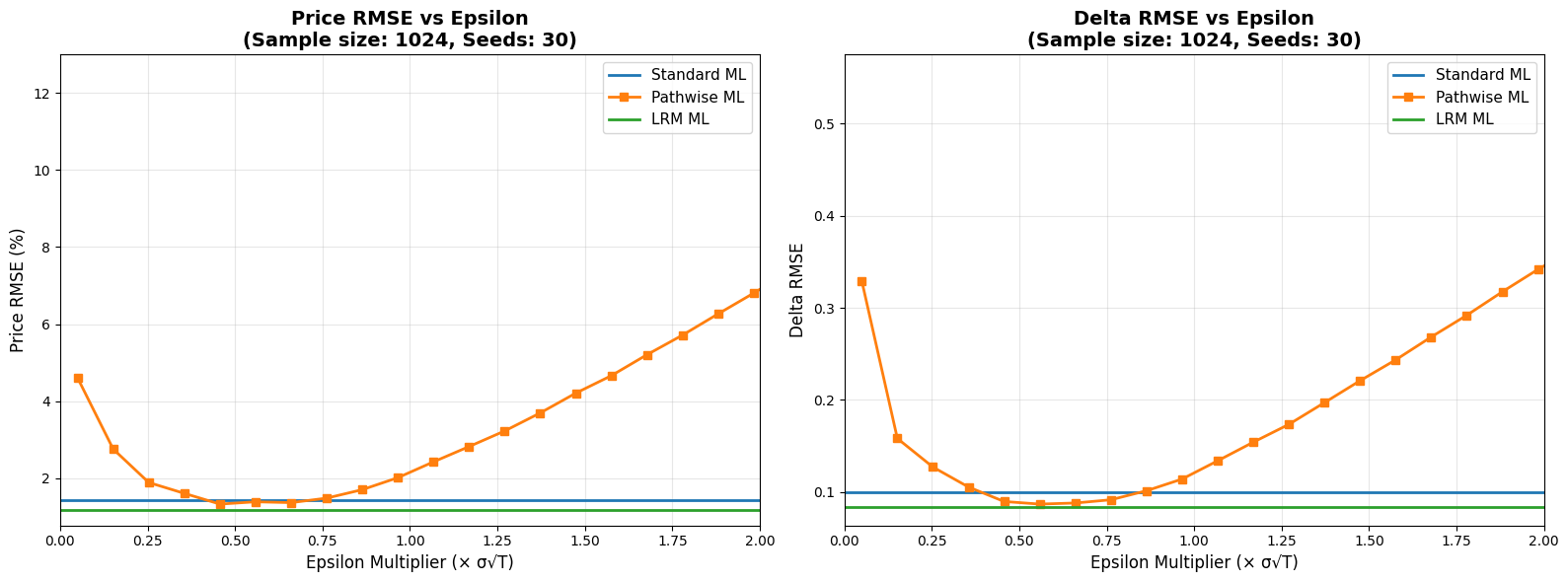}
    \caption{Performance of pathwise ML on a digital option with smoothing is sensitive to smoothing parameter $\epsilon$. Results are averaged over 30 replications, each using a training sample size of 1024.}
    \label{f:epsilon}
\end{figure}

 Figure~\ref{f:epsilon} 
 plots the price RMSE as a percentage of price for a digital option at various levels of $\epsilon$, shown as multiples of $\sigma\sqrt{T}$. The performance of DML with pathwise derivatives is very sensitive to the choice of $\epsilon$. At most values of $\epsilon$, it underperforms standard ML; at all values, it underperforms DML with LRM labels.

 The digital payoff is the simplest case of a discontinuous payoff, yet even here the effective use of smoothing is not obvious. Proper smoothing is even less obvious when discontinuities are path-dependent or dependent on multiple assets.
 Moreover, Price RMSE and Delta RMSE may be minimized at different degrees of smoothing.

\subsection{Barrier Options}
\label{s:barrier}

Barrier options introduce a discontinuity that depends on the path of the underlying, rather than on the terminal value alone.  
We consider a down-and-out call option where the barrier is monitored at intermediate times $T_1,\dots,T_{n-1}$ and the option expires at $T_n$.  
The payoff is given by
\begin{equation}
\pi(S_{T_1},\dots,S_{T_n}) = e^{-rT_n} \mathds{1}_{\{M_n> B\}} (S_{T_n} - K)^+,
\label{pibarrier}
\end{equation}
with $M_n = \min\{S_{T_1},\dots,S_{T_{n-1}}\}$,
so that the contract is knocked out if the underlying is below the barrier $B$
at any time $T_1,\dots,T_{n-1}$.
In the Black-Scholes setting, the underlying is simulated using
\begin{equation}
S_{T_{i+1}} = S_{T_i} \exp\!\Big((r-\tfrac{1}{2}\sigma^2)(T_{i+1}-T_i) + \sigma \sqrt{T_{i+1}-T_i}\, \xi_{i+1}\Big), \quad \xi_{i+1} \sim \mathcal{N}(0,1),
\end{equation}
for $i=0,1,\dots$, with $T_0=0$ and $S_0=x$.
The pathwise derivative of (\ref{pibarrier}) is given by
$$
\partial_x\pi(S_{T_1},\dots,S_{T_n}) = e^{-rT_n} \mathds{1}_{\{M_n> B, S_{T_n}>K\}} S_{T_n}/x.
$$
It reflects the sensitivity of the payoff to $S_{T_n}$, but it fails to capture a potential change in $\mathds{1}_{\{M_n> B\}}$ and is therefore biased.
In contrast, the LRM estimator
\begin{equation}
\widehat{\Delta}_{\text{LRM}}
= \pi(S_{T_1},\dots,S_{T_n}) \frac{\xi_1}{x\sigma\,\sqrt{T_1}},
\label{barrierlrm}
\end{equation}
is unbiased. The LRM factor in (\ref{barrierlrm}) does not depend on 
$\xi_2,\dots,\xi_n$ or $T_2,\dots,T_n$ because, given $S_{T_1}$, the distribution of 
$(S_{T_2},\dots,S_{T_n})$ does not depend on the spot price $x$.

\begin{figure}[H]
    \centering
\includegraphics[trim=0.9in 3.7in 0in 0.4in,clip,width=0.95\textwidth]{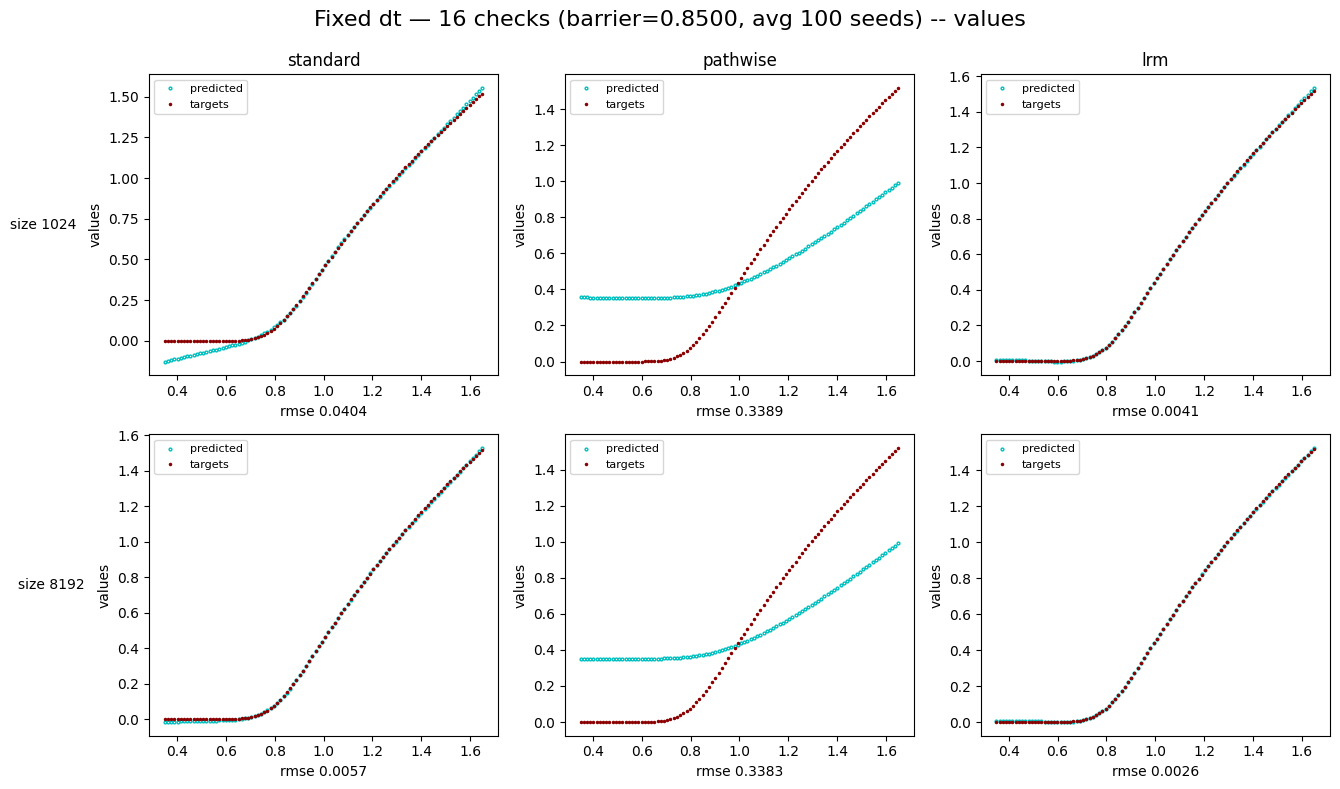}
    \caption{Barrier options — predicted vs. target prices under standard, pathwise, and LRM training, each using a training sample size of 1024, with predictions averaged over 10 replications.  RMSEs are computed from averaged predictions.} 
    \label{fig:barrier_prices}
\end{figure}

\begin{figure}[H]
    \centering
    \includegraphics[trim=0.9in 3.7in 0in 0.4in,clip,width=0.95\textwidth]{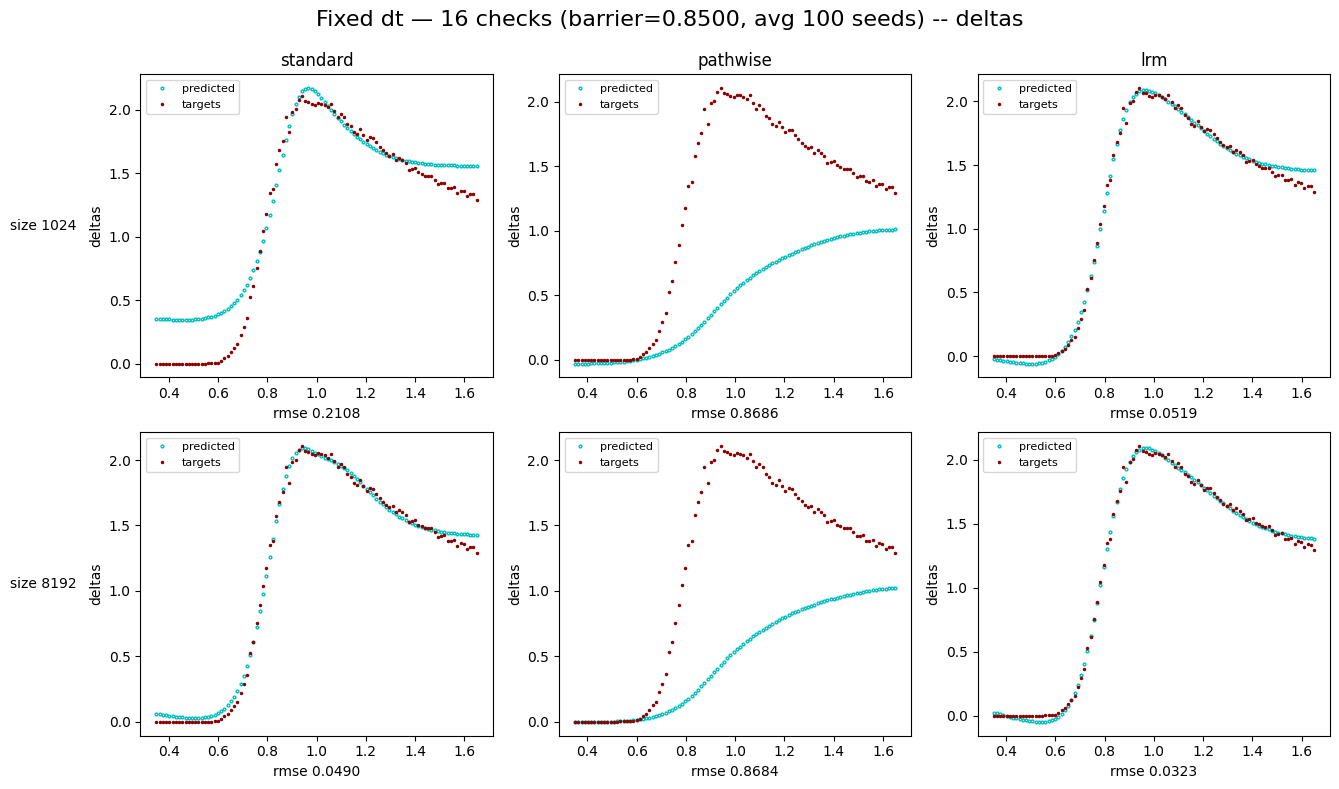}
    \caption{Barrier option deltas — predicted vs. target sensitivities under standard, pathwise, and LRM training, each using a training sample size of 1024, with predictions averaged over 10 replications.  RMSEs are computed from averaged predictions.}
    \label{fig:barrier_deltas}
\end{figure}

Figure~\ref{fig:barrier_prices} compares predicted and target prices under the three training regimes, averaged over 10 independent replications, with the barrier at 85 percent of the spot price, $n=16$ equally spaced dates $T_{i+1}-T_i = 1/3$, 
and $k=100$ in (\ref{kavg}).
Differential ML with pathwise labels performs poorly and shows significant bias. LRM achieves an 83-fold reduction in Price RMSE compared with the pathwise method, and a 10-fold improvement over standard ML.

Figure~\ref{fig:barrier_deltas} makes a similar comparison for barrier option deltas. 
The standard and pathwise trained methods produce deltas with large errors.
The LRM trained method produces much more accurate deltas.

We have found that the results are largely insensitive to the number of steps $n$
and may even improve at large $n$; see Table~\ref{t:barrier}.
We noted in Section~\ref{s:variance} that the variance of LRM estimators sometimes
grows with the number of time steps. But that phenomonen applies when each step
contributes to the score function. In estimating delta, as in (\ref{barrierlrm}), only the
first state transition contributes to the score, so the variance does not grow with
the number of steps.

\begin{table}
\centering
\begin{tabular}{rcccc}
$n=$ &		2	&	4	&	8	&	16 \\ \hline
Price RMSE	&	0.022	&	0.020	&	0.017	&	0.015 \\
Delta RMSE	&	0.144	&	0.137	&	0.114	&	0.104 \\ \hline
\end{tabular}
\caption{LRM accuracy for $n$-step barrier option at different values of $n$
with sample size $m=1024$.}
\label{t:barrier}
\end{table}

\subsection{Discretized Processes}

Our examples  have relied on the availability of explicit transition densities in the Black-Scholes and Bachelier models. We briefly point out how the scope of LRM extends to more general models.

Consider a model defined by a possibly vector-valued process $X_t$ specified through a stochastic differential equation
\begin{equation}
dX_t = \mu(X_t)\, dt + \sigma^{\top}(X_t)\, dW_t.
\label{xt}
\end{equation}
In general, the density of $X_t$ is unknown. However, such a process is typically simulated through an Euler scheme on a time grid $\{0,\Delta t, 2\Delta t,\dots\}$,
with $X_0=x$, and
\begin{equation}
X_{i+1} = X_i + \mu(X_i)\Delta t + \sigma^{\top}(X_i)\sqrt{\Delta t} \xi_{i+1},
\quad \xi_{i+1} \sim \mathcal{N}(0,I).
\label{euler}
\end{equation}
For this discrete-time approximation, the conditional density $p(X_{i+1}|X_i)$ of $X_{i+1}$ given $X_i$ is Gaussian with mean $X_i + \mu(X_i)\Delta t$ and covariance $\sigma(X_i)\sigma^{\top}(X_i)\Delta t$. The density of the path $(x,X_1,\dots,X_n)$ is the product  $p(X_1|x)\cdots p(X_n|X_{n-1})$ of these Gaussian densities. The LRM score function can then be written explicitly by differentiating the log of this product.
(See Section 7.3.4 of Glasserman \cite{pgbook} for additional details.) 
The resulting LRM estimator is unbiased for the Euler scheme: it does not introduce any discretization error beyond that introduced by the Euler scheme itself.
Similar ideas can be applied to other time-discretizations of complicated models.


To illustrate, we consider an Euler discretization of a Heston model
\begin{eqnarray*}
S_{i+1} &=& S_i \exp(-(V^+_i/2)\Delta t + \sqrt{V^+_i\Delta t}(\rho \xi_{1,i+1}+\sqrt{1-\rho^2}\xi_{2,i+1})) \\
V_{i+1} &=& V_i + \kappa(\theta - V_i^+)\Delta t + \eta\sqrt{V_i^+}\sqrt{\Delta t}\xi_{1,i+1},
\end{eqnarray*}
where $\xi_{1,i+1}$ and $\xi_{2,i+1}$ are independent standard normal random
variables. We take $S_0=1$, $\kappa=1$, $V_0=\theta=0.04$, $\eta=0.15$, and
$\rho = -0.7$. We consider a digital call struck at $K=1$ that expires in 10 steps and vary $\Delta t$. The score function is given by $\xi_{2,1}/S_0\sqrt{(1-\rho^2)V_0\Delta t}$.
We evaluate performance in comparison to values estimated from 10 million Monte Carlo paths. Our DML results use $k=100$ in (\ref{kavg}) and a sample size of $m=1024$.

The pathwise method is inapplicable to this problem because the payoff is discontinuous. Table~\ref{t:heston} compares price and delta RMSEs with and without LRM labels. With larger time steps, LRM outperforms the standard method, which does not use derivative labels. At smaller $\Delta t$, the variance of the score function increases and LRM underperforms.

\begin{table}
\centering
\begin{tabular}{c|cccc}
   &	\multicolumn{2}{c}{Price RMSE}	& \multicolumn{2}{c}{Delta RMSE} \\ \hline
$\Delta t$	& Standard	& LRM&	Standard&	LRM \\
0.10&	0.006&	0.008&	0.058&	0.067 \\
0.25&	0.007&	0.006&	0.053&	0.046 \\
0.50&	0.011&	0.005&	0.096&	0.034 \\ \hline
\end{tabular}
\caption{Results for discretized Heston model}
\label{t:heston}
\end{table}

In general, with small $\Delta t$, the LRM estimator based on the Euler scheme (\ref{euler}) may have large variance. Chen and Glasserman \cite{cg} show that more stable estimates can be obtained by averaging multiple estimates to which all the $\xi_{i+1}$ in (\ref{euler}) contribute. They show that  in the limit as $\Delta t\to 0$, this approach recovers Malliavin derivatives calculated from the continuous-time process (\ref{xt}).
Malliavin derivatives, as proposed in Fourni\'{e} et al. \cite{malliavin}, can also be used directly as differential labels, although they are often more difficult to apply.

\section{Gamma}
\label{s:gamma}

We now consider further augmenting the loss function to an expression of the form
\begin{equation}
\mathcal{L}_{\text{Gamma}}(w)
= \frac{1}{m}\sum_{i=1}^{m} \Big[
(\hat{f}^{(i)} - f^{(i)})^2
+ \lambda_1 \|\nabla_x\hat{f}^{(i)} - \delta^{(i)}\|^2
+ \lambda_2\|\nabla_x^2 \hat{f}^{(i)} - \gamma^{(i)}\|^2
\Big].
\label{eq:gamma_loss}
\end{equation}
The second derivatives $\nabla_x^2\hat{f}$ of the neural network approximation are easy to evaluate using the AAD recursion (\ref{aad1})--(\ref{aad3}) with the output $z_L$ set equal to $\nabla_x \hat{f}$ rather than $\hat{f}$. The challenge lies in finding unbiased gamma labels $\gamma^{(i)}$.

The pathwise method virtually never provides unbiased gamma estimates. The problem can be seen through the case of a standard call payoff in the Black-Scholes model. Differenting pathwise once yields the pathwise delta in (\ref{pwcall}), which is discontinuous in the underlying. A second application of pathwise differentiation would therefore produce biased gamma estimates.

Following the steps leading to (\ref{pilrm}) yields the following LRM estimator
of gamma,
$$
\pi(S_T)\frac{\nabla^2_x p(S_T;x)}{p(S_T;x)},
$$
which is broadly applicable. For payoffs without discontinuities, we can also use a hybrid pathwise-LRM estimator, which first differentiates pathwise and then applies LRM to differentiate a second time. In the case of a standard call in the Black-Scholes model, differentiating pathwise once yields (\ref{pwcall}). 
The resulting expression has an explicit dependence on $x$, which can be differentiated directly, and an implicit dependence on $x$ through the distribution of $S_T$, which can captured by applying LRM. The combination yields
$$
\widehat{\Gamma}_{\text{PW--LR}}
=\partial_x\delta + \delta\nabla_xp(S_T;x)
= \mathds{1}\{S_T>K\}\frac{e^{-rT}S_T}{x^2}\left(\frac{\xi}{\sigma\sqrt{T}}-1\right).
$$

\begin{figure}
\centerline{
\includegraphics[trim=10in 4.8in 0in 0.8in,clip,width=2.5in]{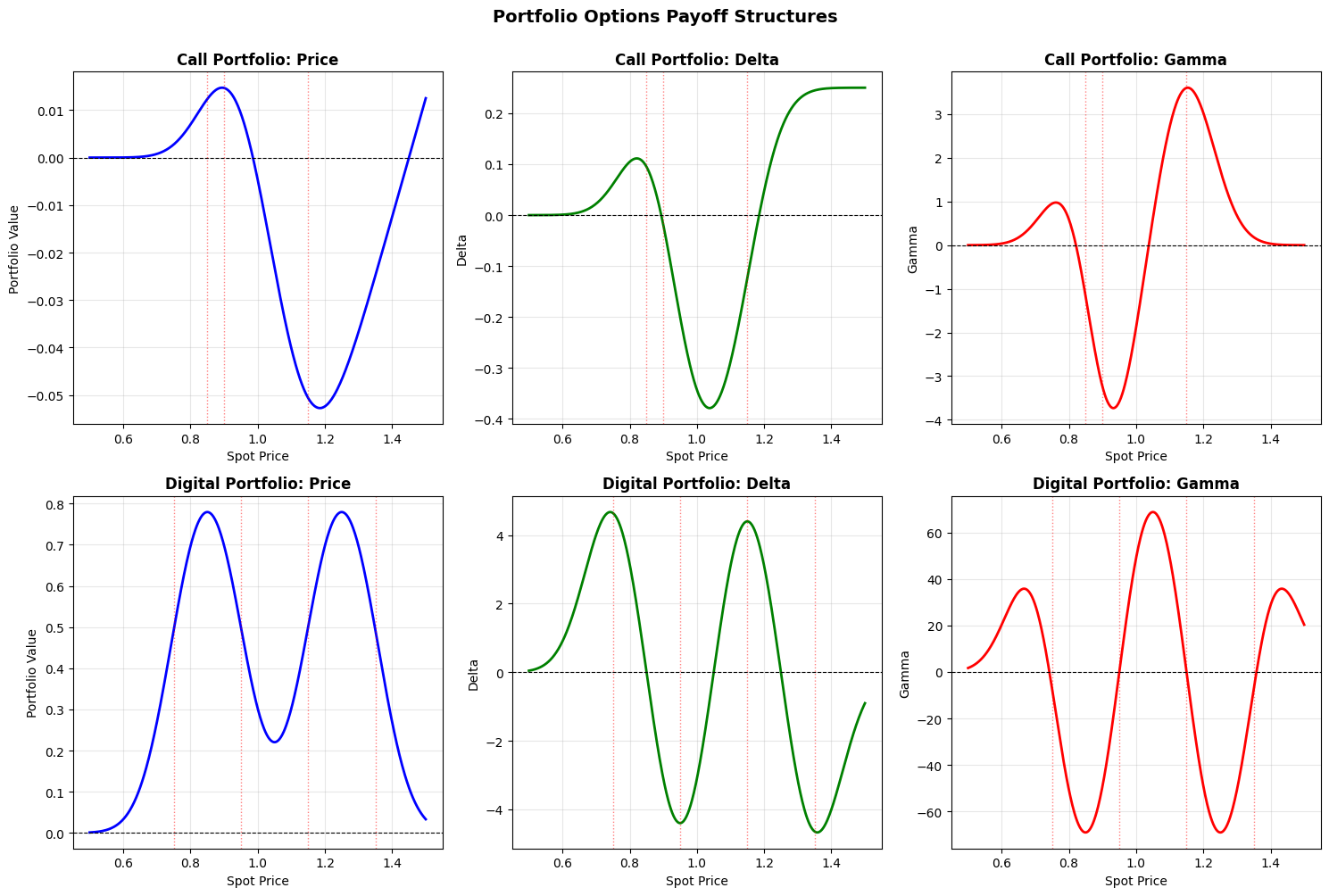}
\quad
\includegraphics[trim=11.8in 0in 0in 0.24in,clip,width=3.0in]{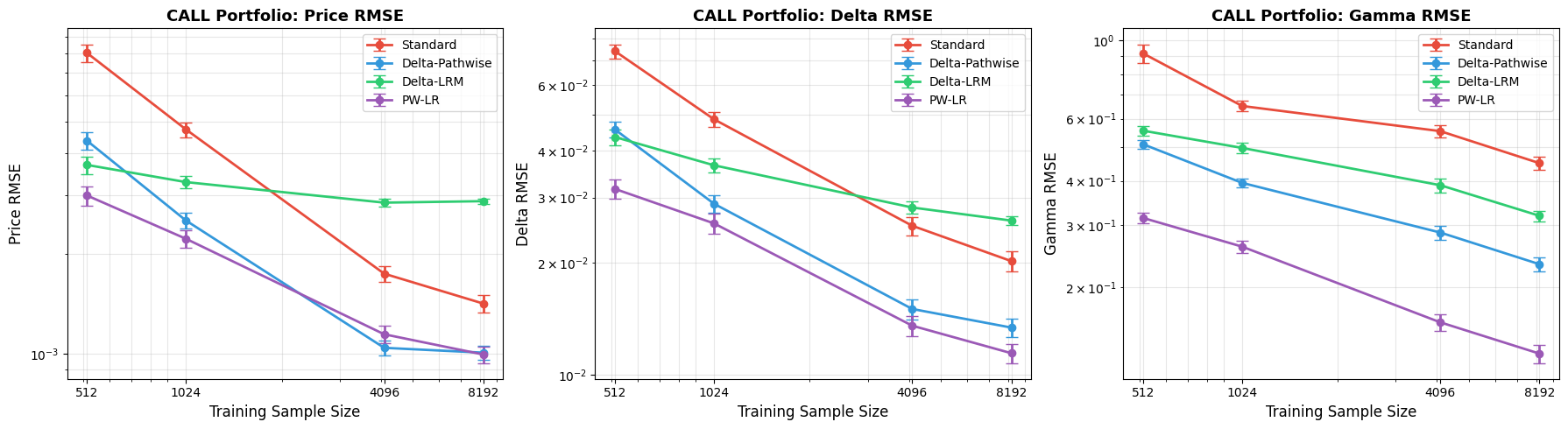}
}
\caption{Left: Gamma of the option portfolio. Right: RMSEs for gamma estimates. The derivative-regularized methods produce smaller errors than standard ML, and PW-LR achieves the smallest errors}
\label{f:gamma}
\end{figure}

To illustrate, we consider an options portfolio with payoff
$$
(S_T-0.85)^+  - 1.5(S_T-0.9)^+ + 0.75(S_T-1.15)^+.
$$
We take  $r=0$, $T=1/3$, and $\sigma=0.20$. 
The left panel of Figure~\ref{f:gamma} plots the gamma for this portfolio as a function of the spot price $S_0$. The right panel compares RMSEs for gamma estimates using four methods: standard ML (trained only on payouts), Delta-Pathwise and Delta-LRM (trained on payouts and delta labels), and PW-LR, trained on payouts, pathwise delta labels, and PW-LR gamma labels. The pathwise delta labels are unbiased for this problem because the payoffs are continuous.
The two delta-supervised methods give equal weight to the payout and delta components of the loss function (\ref{eq:DMLloss}). The PW-LR methods gives 20 percent weight to the gamma component of (\ref{eq:gamma_loss}) and 40 percent weight to each of the other two components.
The right panel of Figure~\ref{f:gamma} shows that the PW-LR method performs best overall, and is consistent as sample sizes increase.

\section{Conclusions}

Differential ML is a valuable tool for improving the training of neural network approximations to complex simulation models used for pricing and risk management. It works by targeting price sensitivities as well as prices in the model training process. 
It has two potential benefits: regularizing the fit to prices, and improving the accuracy of derivatives calculated from a neural network approximation.

To achieve these benefits, Differential ML requires unbiased differential labels. Pathwise derivatives, used in Huge and Savine \cite{hs}, are biased for discontinuous payoffs, such as those with barrier or digital features. Our examples show that this bias can severely degrade the accuracy of this method. We have shown that in these cases, better performance is attained using differential labels calculated through the likelihood ratio method. A further extension accommodates gamma labels in the training process. The Differential ML framework is sufficiently flexible to incorporate these alternative labels with only minor changes to the implementation.
The main limitation of the LRM method is that it often produces large variance when the simulation time step is very short.
\bigskip

{\noindent\bf Acknowledgements.}  The authors thank Brian Huge, Antoine Savine, and an anonymous reviewer for helpful comments.


\end{document}